\documentstyle[aps,prl,epsfig,twocolumn]{revtex}
\begin{document}

\draft

\title{Phase Transitions in a Two-Component Site-Bond Percolation Model}

\author{H. M. Harreis and W. Bauer} \address{National Superconducting
  Cyclotron Laboratory and Department of Physics and Astronomy,
  Michigan State University, East Lansing, Michigan 48824}

\maketitle

\begin{abstract}
  A method to treat a $N$-component percolation model as effective one
  component model is presented by introducing a scaled control
  variable $p_{+}$. In Monte Carlo simulations on $16^{3}$, $32^{3}$,
  $64^{3}$ and $128^{3}$ simple cubic lattices the percolation
  threshold in terms of $p_{+}$ is determined for $N=2$.  Phase
  transitions are reported in two limits for the bond existence
  probabilities $p_{=}$ and $p_{\neq}$. In the same limits, empirical
  formulas for the percolation threshold $p_{+}^{c}$ as function of
  one component-concentration, $f_{b}$, are proposed. In the limit
  $p_{=} = 0$ a new site percolation threshold, $f_{b}^{c} \simeq
  0.145$, is reported.
\end{abstract}


\narrowtext

The percolation model goes back to Flory \cite{Flory1} who introduced
it in the context of polymer gelation. Since then it has been used in
a wide range of approaches and
techniques\cite{SykesEssam,Bauerall1,GalamMauger1,vanderMarck1,GalamMauger2,Bauerall2,Sahimibook,ContBouchaud}.
In standard percolation models either bond or site percolation is
dealt with\cite{Staufferbook,Grimmettbook}. Site-bond percolation
\cite{ConiglioStanleyKlein,HeermannStauffer} combines the two
formulations, dealing with randomly occupied sites (vertices) and
randomly existing bonds (open edges) connecting these sites. However
in this version of the model only one active component exists, the
other sites are considered unoccupied. A further generalization is to
consider several components, which was done for site percolation as
well as bond percolation by Zallen \cite{Zallen1} and called
polychromatic percolation. Zallen focused on the coexistence of
percolating species in highly connected lattices, giving a criterion
for the occurrence of a panchromatic regime where all species
percolate. Site-bond percolation using two components was investigated
previously by one of us \cite{Bauerall3} and applied to the question
of the nuclear liquid gas phase transition. Site-bond percolation with
several species was considered in \cite{Ioselevich} and an approximate
percolation criterion was given.

In this Letter, we investigate a two component site-bond percolation
model on a simple cubic lattice, focusing on two specific limits which
exhibit novel behavior. Let us begin by describing the approach we
have taken, in the general case of $N$ different component flavors. No
assumption concerning topological dimensions or lattice structure is
made. We have $N$ component concentrations $f_{i}$ with $\sum_{i =
  1}^{N} f_{i} = 1$ and different bond probabilities to connect all
possible combinations of sites, resulting in $A_{par} = (N - 1) + {N +
  2 - 1 \choose 2}$ free parameters $a_{i}$. The bonds have been
assumed to be directionless, meaning that their probabilities only
depend on the species of the sites they are connecting. We now want to
know in which region of this $A_{par}$-dimensional parameter space an
infinite network $C_{\infty}$ of connected bonds occurs, that is,
where the probability for a given site to belong to the infinite
network, $p_{\infty}(\{a_{i}\})$, is non-zero. The particular type of
a bond shall be irrelevant in order for it to belong to the infinite
network.  For a system with $N \geq 3$ components, however, this
approach is quite impractical. It would be preferable to be able to
reduce the dependence of the order parameter, $p_{\infty}$, to one
variable at fixed particle concentrations. We propose one such
variable in following the definitions of \cite{Bauerall3},
generalizing them to $N$ components. In analogy to the bond existence
probability used in ordinary one component bond percolation models,
which gives the bond density in the system under observation, we
introduce the scaled control parameter $p_{+}$:
\begin{equation}
  \label{eq:pplusgeneral}
  p_{+} = \sum_{i \geq j = 1}^{N} \alpha_{i j} p_{i j}.
\end{equation}
Here, the $p_{i j}$'s denote the probability for a bond to exist
between two sites occupied by species $i\ \text{and}\ j$ and
$\alpha_{i j} = \alpha_{j i}$ is the probability that any given
nearest neighbor edge is one that connects two sites of flavors $i\ 
\text{and}\ j$:
\begin{equation}
  \label{eq:alphaij}
  \alpha_{i j} = 2 f_{i} f_{j},
\end{equation}
with the constraint that $\sum_{i \geq j}^{N} \alpha_{i j} = 1$.

\begin{center}
\begin{figure}[hbt]
  \epsfig{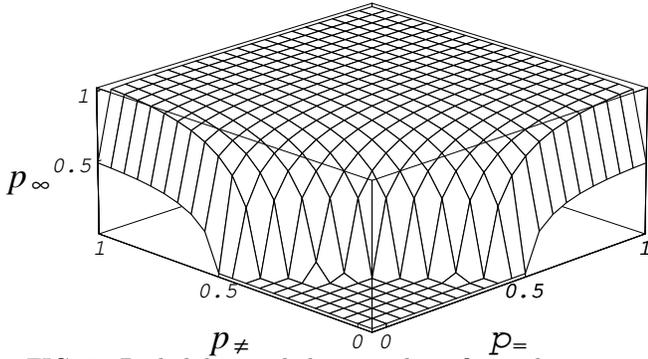} 
\caption{Probability to belong to the infinite cluster, $p_{\infty}$, 
  in a simple cubic two component site-bond percolation lattice of
  size $128^{3}$ as a function of the two parameters, $p_{=}\ 
  \text{and}\ p_{\neq}$, calculated for a fraction of the blue species
  of $f_{b} = 0.5$.}
  \label{fig:3dpinf}
\end{figure}
\end{center}

\begin{center}
\begin{figure}[hbt]
  \epsfig{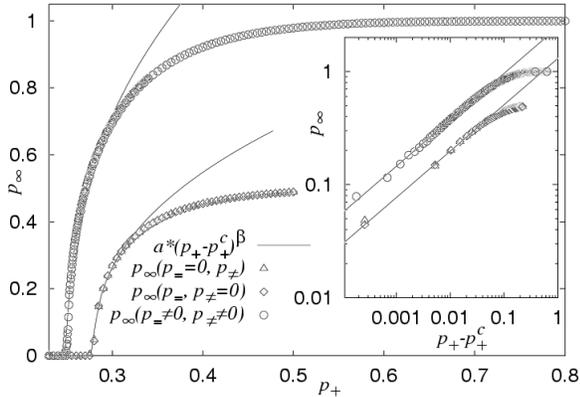} 
\caption{Probability to belong to the infinite cluster, $p_{\infty}$,
  as a function of the scaled control parameter, $p_{+}$, in a
  $128^{3}$ simple cubic lattice with $f_{b} = 0.5$, for values of
  $p_{=}, p_{\neq} \in [0,1]$, fitted with $p_{\infty} \propto (p_{+}
  - p_{+}^{c})^{\beta}$ (solid line). Here $p_{+}^{c} \simeq 0.251$
  for the 'upper' branch, $p_{+}^{c} \simeq 0.280$ for the 'lower'
  branch and $\beta=0.41$ in both cases. The inset shows the same data
  in a double logarithmic representation.}
  \label{fig:pinfpplusall}
\end{figure}
\end{center}

For our simulation we consider a two component system on $16^{3}$,
$32^{3}$, $64^{3}$ and $128^{3}$ simple cubic lattices. For simplicity
we shall call one species blue, the other red. We now have four free
parameters to vary: The fraction of one of the components, say of the
blue sites, $f_{b}$ and three bond activation probabilities: $p_{bb}$
for bonds connecting two blue sites, $p_{rr}$ for bonds connecting two
red sites and $p_{\neq}$ for $b\text{-}r$-bonds. However we shall set
$p_{=} \equiv p_{bb} = p_{rr}$, introducing a symmetry in the system.
This is motivated by considerations of, for example, isospin symmetry,
where the $e^{+}e^{+}$ and $e^{-}e^{-}$ interactions are identical. Eqs.
(\ref{eq:pplusgeneral}) and (\ref{eq:alphaij}) now read
\begin{equation}
  p_{+} = \alpha_{\neq}\ p_{\neq} + \alpha_{=}\ p_{=},
  \label{eq:pplustwo} 
\end{equation}
with $\alpha_{=} = (1-\alpha_{\neq})$, and
\begin{equation}
  \alpha_{\neq} = 2 f_{b} (1-f_{b}),
  \label{eq:alphaneq}
\end{equation}
respectively, where we have replaced the double indices by a more
intuitive notation for only two components. Again the question is in
which region of the three dimensional $p_{=} \text{---} p_{\neq}
\text{---} f_{b}$ space an infinite network of bonds appears in the
lattice. We defer the question of the concentration dependence to
later and for the moment set $f_{b}$ to some fixed value, which shall,
for now, be $f_{b}=0.5$. In the simulation the lattice is populated at
random, without correlations, according to $f_{b}$ and bonds are
formed for varied values of $p_{=}, p_{\neq} \in [0,1]$ using a Monte
Carlo algorithm. The resulting cluster structure is analyzed using a
cluster-find-algorithm described in \cite{Bauerall1} and
$p_{\infty}(p_{=},p_{\neq})$ is recorded. As always, a cluster is
defined as a set of vertices connected by open edges. In Fig.
\ref{fig:3dpinf} we show $p_{\infty}$ as a function of the two control
parameters, $p_{=}\ \text{and}\ p_{\neq}$, at a concentration of
$f_{b} = 0.5$. We can see that $p_{\infty}$ changes from 0 (front
corner) to 1 (back corner), with a critical line of a second order
phase transition in the $(p_{=},p_{\neq})$-plane. We now follow our
previous consideration and analyze the same data in terms of the
scaled control parameter $p_{+}$, the result of which is displayed in
Fig.  \ref{fig:pinfpplusall}. Two distinct branches are seen, both in
the shape of a second order phase transition. Figure
\ref{fig:pinfpplusall} first suggests that $p_{+}$ is a good control
parameter, as we only have these two universal scaled curves, but
furthermore that another transition seems to take place in the system.
Analysis of the data shows that the 'upper' branch constitutes of
points with both $p_{=}$ and $p_{\neq}$ non-zero, whereas all points
with $(p_{=} = 0,p_{\neq} \neq 0)$ and $(p_{=} \neq 0,p_{\neq} = 0)$
fall on the 'lower' branch. For other values of $f_{b}$ the same
behavior is found, but then the two curves for the zero-limits in
$p_{=}$ or $p_{\neq}$ are not the same. Also shown in Fig.
\ref{fig:pinfpplusall} are the expectations from one component bond
percolation theory, $p_{\infty} \propto (p_{+} - p_{+}^{c})^{\beta}$,
(smooth lines). For the finite $(p_{=},p_{\neq})$ regime the critical
value $p_{+}^{c}$ has the same numerical value as the bond existence
probability in one component bond percolation (aside from a small
difference due to finite size effects), $p_{+}^{c} = 0.251 \pm
0.002$. As long as both bond-types are active, the system under
observation here and the one component model show an identical phase
transition behavior, which is consistent with the findings presented
in \cite{Bauerall3}. In the zero-limits of $p_{=}$ or $p_{\neq}$,
however, $p_{+}^{c}$ is shifted to $p_{+}^{c} = 0.280 \pm 0.002$.
The critical exponent $\beta = 0.41$ from one component bond
percolation theory is the same in both cases presented here, as shown
in the double logarithmic plot in the inset of Fig.
\ref{fig:pinfpplusall}.

\begin{center}
\begin{figure}[hbt]
  \epsfig{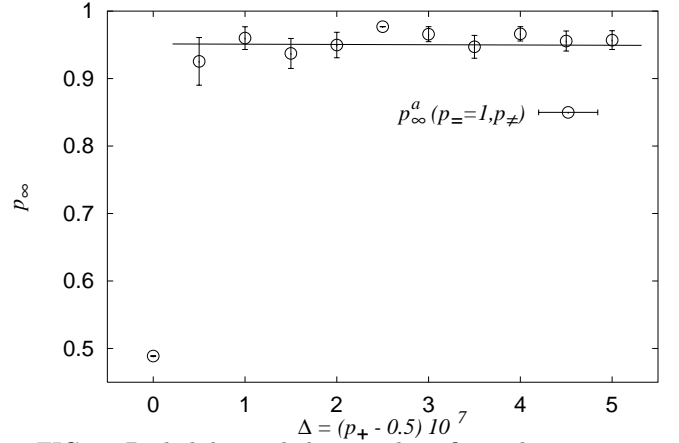}
\caption{Probability to belong to the infinite cluster, $p_{\infty}$,
  as a function of the scaled control parameter $p_{+}$, in a
  $128^{3}$ simple cubic lattice with $f_{b} = 0.5$, at fixed $p_{=} =
  1$ for varied $p_{\neq}$. $50$ independent simulation events where
  taken into account and the average was taken over events in which
  the number of $br$-bonds actually formed in the simulation,
  $n_{br}$, was non-zero.}
  \label{fig:pinfpeq1}
\end{figure}
\end{center}

\begin{center}
\begin{figure}[hbt]
  \epsfig{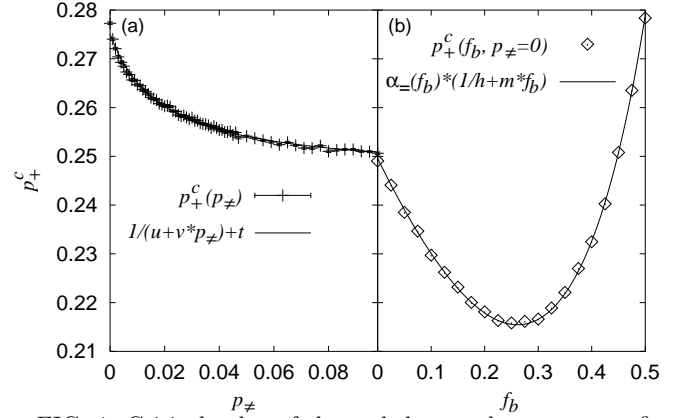} 
\caption{Critical value of the scaled control parameter, $p_{+}^{c}$,
  in a $128^{3}$ simple cubic lattice with $f_{b} = 0.5$, (a) plotted
  as a function of $p_{\neq}$, and (b) plotted as a function of the
  fraction of blue sites, $f_{b}$, in the limit $p_{\neq} =0$ and
  obtained by a fit to the scaling law $|p_{+}^{c}(L) - p_{+}^{c}|
  \propto L^{-1/\nu}$. The errors, estimated as described in
  \protect\cite{vanderMarck1}, are smaller than the symbol sizes in
  (b).}
  \label{fig:ppluscvspneqandppluscvsfbpneq0}
\end{figure}
\end{center}

What causes the change of $p_{+}^{c}$ in the limits $p_{=} \to 0$ and
$p_{\neq} \to 0$? We restrict ourselves to a discussion of $p_{\neq}
\to 0$, as the same line of arguments applies in the other limit. For
simplicity we first choose $p_{=} = 1$. Then setting $p_{\neq} = 0$
corresponds to a lattice in which all available $bb$- and $rr$-edges
are open, but all $br$-edges are closed. There will be an infinite
network present, as usual identified with the biggest percolating
cluster (with two components each above the site percolation threshold
being present we could have two percolating clusters), with
$p_{\infty}$ given by
\begin{equation}
  \label{eq:pinftylimpneq0}
  p_{\infty}(p_{\neq}=0) = f_{b} - \sum_{s=1}^{\infty -1} n_{s}^{b}\ s,
\end{equation}
where $n_{s}^{b}$ is the number of $b$-clusters of size $s$ and the
upper limit in the sum is meant to indicate that the infinite cluster
is excluded. Letting $p_{\neq}=\epsilon=1/N_{edges}$ introduces on
average one $br$-bond in the lattice. This bond might, with some small
probability, which is related to the number of perimeter edges of all
finite clusters, connect two finite clusters. With a noticeably higher
probability however, it will open an edge that connects the infinite
cluster to the biggest $r$-cluster, $C_{max}^{r}$, with $|C_{max}^{r}|
\approx |C_{\infty}|$. The infinite cluster can now consist of both
$b$- and $r$-sites and $p_{\infty}$ reads
\begin{equation}
  \label{eq:pinfty}
    p_{\infty}(p_{\neq}=\epsilon) = 1 - \sum_{s=1}^{\infty -1} n_{s}\ s, 
\end{equation}
which, by rewriting $f_{b}$ in Eq. \ref{eq:pinftylimpneq0} as $f_{b} =
1-\sum_{s=1}^{max} n_{s}^{r}\ s$, leads to a difference in
$p_{\infty}$:
\begin{eqnarray}
  \label{eq:deltainf}
  \delta_{\infty}(\epsilon) &=
  &p_{\infty}(p_{\neq}=\epsilon)-p_{\infty}(p_{\neq}=0)\nonumber\\
  &= &\Bigl(\sum_{s=1}^{\infty -1}
  n_{s}^{b}\ s + \sum_{s=1}^{max} n_{s}^{r}\ s\Bigr) -
  \sum_{s=1}^{\infty -1} n_{s}\ s,
\end{eqnarray}
with the property that $\lim_{\epsilon \to 0}
\delta_{\infty}(\epsilon) \neq 0$ for all events where $C_{\infty}$
and $C_{max}^{r}$ are connected. Averaging over all events will yield
some effective $\delta_{\infty}^{\text{\it{eff}}}$ with $0 <
\delta_{\infty}^{\text{\it{eff}}} < \delta_{\infty}$. This leads to
the conclusion that in the limit $p_{\neq} \to 0$ a first order phase
transition takes place in $p_{\infty}$. The simulation results support
this conjecture. Figure \ref{fig:pinfpeq1} shows these results, where
the average was taken over events in which the number of $br$-bonds
actually formed in the simulation, $n_{br}$, was non-zero. For values
of $p_{=}$ other than $1$ one still finds the same behavior, somewhat
less pronounced due to smaller $|C_{\infty}|$ and $|C_{max}^{r}|$. The
same holds for $f_{b} \neq 0.5$ where it is clear that the effect
vanishes continuously in the limit of a one component bond percolation
system: $f_{b} \to 0$ or $f_{b} \to 1$. Obviously the question arises
what the behavior of $p_{+}^{c}$ is. We found that for $p_{\neq} \to
0$ the percolation threshold $p_{+}^{c}$ continually changes to a new,
concentration dependent value. In Fig.
\ref{fig:ppluscvspneqandppluscvsfbpneq0}(a) we show this transition,
which is of second order type. We also show a fit to the curve, which
is of the form
\begin{equation}
  \label{eq:ppluscvspneq}
  p_{+}^{c}(p_{=}, p_{\neq} \to 0)= \frac{1}{(u + v\ p_{\neq})} +t,
\end{equation}
where the fit parameters were found to be $u=34.6 \pm 0.3$, $v=1823
\pm 72$ and $t=0.246 \pm 0.001$. It has to be noted that this
formula only stands on empirical grounds, fits by exponential
functions may also be useful. As stated earlier, the same line of
thought is applicable to the $p_{=} \to 0$ limit and indeed do our
simulations present the same results in the system with $f_{b} = 0.5$.
For $f_{b} \neq 0.5$ this direct symmetry is broken, but qualitatively
the results are still the same. Even for $f_{b}=0.5$ they differ
however with respect to the afore-mentioned dependence of $p_{+}^{c}$
on the concentration $f_{b}$ in the limits $p_{=} = 0$ and $p_{\neq} =
0$. For $p_{+}^{c}(f_{b},p_{\neq} = 0)$ we find the functional form
depicted in Fig.  \ref{fig:ppluscvspneqandppluscvsfbpneq0}(b). The
results shown are fits of $p_{+}^{c}$ to the scaling relation,
\begin{equation}
  \label{eq:scaling}
  |p_{+}^{c}(L) - p_{+}^{c}| \propto L^{-1/\nu},
\end{equation}
as given in Ref. \cite{Staufferbook}, where we kept $\nu$ fixed at
$0.88$ and lattice sizes $L=16$, $L=32$, $L=64$ and $L=128$ were taken
into account. The simulation data is fitted with
\begin{equation}
  \label{eq:pplusvsfbempirpneq0}
  p_{+}^{c}(f_{b},p_{\neq}=0) = \frac{2 (f_{b} - \frac{1}{2})^{2} +
    \frac{1}{2}}{h + m f_{b}}.  
\end{equation}
This is Eq. (\ref{eq:pplustwo}) with the purely empirical assumption
of a hyperbola for $p_{=}^{c}(f_{b})$, which is in agreement with the
results of Heermann and Stauffer for a one component site-bond model
\cite{HeermannStauffer}. Fitting the parameters to our simulation data
results in $h=4.007 \pm 0.002$ and $m=-4.428 \pm 0.005$. A comparison
with the formula given by Heermann and Stauffer yields $h =
1/p_{bond}^{c}(0) = 4.019$ and $m = (1 - p_{bond}^{c}) /
(p_{+}^{c}(f_{site}^{c}-1)) = - 4.386$, where $p_{bond}^{c} = 0.2488$
and $f_{site}^{c} = 0.3116$ are the percolation thresholds for one
component bond- and site- percolation on a three dimensional simple
cubic lattice, respectively. However, in contrast to the one component
model, due to the symmetry introduced in the system, notably that
$p_{=} \equiv p_{bb} = p_{rr}$, we only have $f_{b} \in [0.0,0.5]$ as
independent regime here, with the interval $f_{b} \in [0.5,1.0]$ being
symmetric to the one shown here, with only the roles of blue and red
sites being switched. This is a manifestation of the two components
behaving like two superposed, non-interfering one component site-bond
percolation systems and lets us conclude that the phase transition in
$p_{+}^{c}$ for $p_{\neq} \to 0$ discussed above may be interpreted as
an effective transition from a one component bond percolation model to
a one component site-bond percolation model. For the case of the
second limit discussed, $p_{=} \to 0$, the results are shown in Fig.
\ref{fig:pneqcandppluscvsfbpeq0}. In Fig.
\ref{fig:pneqcandppluscvsfbpeq0}(a) we go back to the bond existence
probability $p_{\neq}$. Its critical value, as a function of the
concentration, $p_{\neq}^{c}(f_{b})$, is well reproduced with an
exponential fit:
\begin{equation}
  \label{eq:pneqcvsfbempirpneq0}
  \varphi = p_{\neq}^{c}(f_{b}) = a\ \exp(-d\ f_{b})\ + c
\end{equation}
Fitting this empirical formula to the simulation data gives $a = 2.1
\pm 0.07$, $d = 10.9 \pm 0.2$ and $c = 0.547 \pm 0.002$. We define
$f_{b}^{c} \equiv \varphi^{-1}(p_{\neq}\! \! =\! \! 1)$ and numerically get
$f_{b}^{c} \simeq 0.14$ from Eq. (\ref{eq:pneqcvsfbempirpneq0}),
whereas in an independent simulation we find $f_{b}^{c} =0.145 \pm
0.001$, which again is a result from a fit to the scaling relation
Eq. (\ref{eq:scaling}). The parameter $f_{b}^{c}$ can be regarded as a
new threshold, which would correspond to the percolation threshold in
a simple site percolation model in which nearest neighbors only belong
to the same cluster if they are of opposite flavor, unlike the normal
site percolation model which yields $f_{site}^{c}=0.3116$. In Fig.
\ref{fig:pneqcandppluscvsfbpeq0}(b) $f_{b}^{c}$ determines the
critical point of the phase transition line $p_{+}^{c}(f_{b},p_{=} =
0)$. Another argument in favor of a qualitatively new behavior arises
by considering the density of accessible edges at the critical
concentrations in the two models. For the one considered here it is
given by $\alpha_{\neq}(f_{b}^{c})$, see Eq. (\ref{eq:alphaneq}), for
the usual site percolation model we might define one in an analogous
manner: $\alpha_{site} = (f_{site}^{c})^{2}$. By setting
$\alpha_{\neq}(f_{b}^{c}) = \alpha_{site}$ one would expect
$f_{b}^{c}$ to be $0.051$ which stands in contradiction to our
findings.

In concluding, we introduced a new way to treat N-component
percolation. This approach was applied to a two component site-bond
percolation model and new first order phase transitions of
$p_{\infty}$ were reported in the limits $p_{\neq} \to 0$ and $p_{=}
\to 0$. In the latter case we could furthermore establish a novel
empirical formula for the percolation threshold as a function of
component concentration, whereas in the first earlier findings of a
one component site-bond percolation model were found to apply in the
two component model too. The field for future work in this area seems
vast, one might, for example, try to apply the same method to
multi-component systems on lattices of higher dimensions and/or higher
connectivity. This approach should also find a broad range of possible
applications. One might think of special networks or gelation
phenomena with several components involved, which only interact with
each other, as well as wetting phenomena. Furthermore an application
to stock-market simulations seems possible and is being undertaken by
the authors.

This work was supported by the National Science Foundation under Grant
No. PHY-9605207. One of us (H.M.H.) is supported in part by the
Studienstiftung des deutschen Volkes.

\begin{center}
\begin{figure}[hbt]
  \epsfig{file=fig5.epsi, width=3.375in, angle=-90}
\caption{Critical value of $p_{\neq}$ and of the scaled control
  parameter, $p_{\neq}^{c}$ and $p_{+}^{c}$, plotted as a function of
  the fraction of blue sites $f_{b}$ in the limit $p_{\neq} =0$. The
  data has been obtained by a fit to $|p_{+}^{c}(L) - p_{+}^{c}|
  \propto L^{-1/\nu}$. Again the errors, estimated as described in
  \protect\cite{vanderMarck1}, are smaller than the symbol sizes.}
\label{fig:pneqcandppluscvsfbpeq0} 
\end{figure} 

\end{center} 

\end{document}